\documentclass[prints]{aa}
\usepackage{psfig}
\begin{document}

\thesaurus{11.04.2; %Galaxies: dwarf 
           11.06.2; %Galaxies: fundamental parameters
           11.09.1 DDO\,44; %Galaxies: individual 
           11.16.1; %Galaxies: photometry
           11.01.1 %Galaxies: abundances
}

\bigskip
\title{The dwarf spheroidal galaxy DDO 44: stellar populations and 
distance\thanks{Based on observations made with the BTA telescope at the 
Special Astrophysical Observatory operated by the Russian Academy
of Sciences and with the NASA/ESA Hubble Space
Telescope.  The Space Telescope Science Institute is operated by the
Association of Universities for Research in Astronomy, Inc. under NASA
contract NAS 5-26555.}}

\titlerunning{DSph galaxy DDO 44}

\author{I.D.\ Karachentsev \inst{1} \and M.E.\ Sharina \inst{1}
\and E.K.\ Grebel \inst{2}\thanks{Hubble Fellow}
\and A.E.\ Dolphin \inst{2,3} \and
D.\ Geisler \inst{4} \and P.\ Guhathakurta \inst{5}
\and P.W.\ Hodge \inst{2} 
\and V.E.\ Karachentseva \inst{6} 
\and A.\ Sarajedini \inst{7}
\and P.\ Seitzer \inst{8}}

\authorrunning{Karachentsev et al.}

\institute{Special Astrophysical Observatory, Russian Academy
          of Sciences, N.\ Arkhyz, KChR, 357147, Russia
\and Department of Astronomy, University of Washington, Box 351580,
     Seattle, WA 98195, USA
\and  Kitt Peak National Observatory, National Optical Astronomy Observatories,
     P.O. Box 26732, Tucson, AZ 85726, USA
\and  Departamento de F\'{\i}sica, Grupo de Astronom\'{\i}a, Universidad 
     de Concepci\'on, Casilla 160-C, Concepci\'on, Chile
\and  UCO/Lick Observatory, University of California at Santa Cruz, 
     Santa Cruz, CA 95064, USA
\and  Astronomical Observatory of Kiev University, Observatorna 3, 254053, 
     Kiev, Ukraine
\and  Astronomy Department, Wesleyan University, Middletown, CT 06459, USA
\and  Department of Astronomy, University of Michigan, 830 Dennison Building, 
     Ann Arbor, MI 48109, USA}

\maketitle
\begin{abstract}
 We have ground-based and HST WFPC2 imaging of the nearby low surface 
brightness dwarf
spheroidal galaxy DDO~44.  For the first time DDO~44 was
resolved into stars. The
resulting color-magnitude diagram for about 1290 stars show the red giant
branch with a tip at $I = 23.55\pm0.15$, which yields
the distance  $D_{MW}= 3.2\pm0.2$ Mpc consistent with membership of DDO~44
in the NGC~2403 group. The linear separation of DDO~44 from
NGC~2403 is 75 kpc on the sky and $30\pm450$~kpc along the line of sight. The
relationship between the dwarf galaxy's absolute magnitude, 
$M_R^o = -13.1$, the central surface brightness,
$\mu_R(0) = 24.1$ mag arcsec$^{-2}$, and the mean metallicity,
[Fe/H] = $-$1.7 dex follow the trend 
for other nearby dwarf spheroidal galaxies.
One globular cluster candidate has also been identified in DDO~44.
\end{abstract}

\section{Introduction}

In the "Catalogue of low surface brightness galaxies" (Karachentseva~\&
Sharina 1988) there are 1555 objects situated within the Local Supercluster
volume. This sample covering the whole sky shows a striking difference in
distributions of dwarf irregular (dIrr) and dwarf spheroidal (dSph) galaxies 
over the sky (see also Figs.\ 1 and 2 in Karachentseva~\& Vavilova 1994). About
90\% of the dSphs are concentrated towards the  Virgo and Fornax
clusters, while the rest of them are associated with the known nearby
groups of galaxies. In the groups around the Milky~Way, M~31, M~81, and
NGC~5128, which were studied most intensely, the dSph companions
also occupy a smaller volume with $R < 300$ kpc compared to dIrr members
(van den Bergh 1994, Karachentsev 1996).
In contrast, the sky distribution of dIrr galaxies looks rather
smooth: only about half of them are associated with groups and clusters.
Based on these data, one can assume that in the space between clusters and
groups, i.e., in the so-called ``metagalactic field'', spheroidal dwarf systems
may be absent in general. Such a morphological segregation appears quite
natural in an evolutionary picture where irregular dwarf systems are
transformed into spheroidal ones through gas loss by external ram pressure
and tidal stripping as they approach giant galaxies (Einasto et al.\ 1974,
Lin~\& Faber 1983). In this respect the discovery of an
isolated dSph galaxy could shed light on dwarf galaxy evolution.
Da Costa (1994) and Saviane et al.\ (1996) have shown that the Tucana
dwarf galaxy may be considered as an isolated dSph system
located almost at the border of the Local Group at $\sim 0.9$ Mpc from the 
Milky Way.
\begin{figure*}[bt]
%\vbox{
%\special{psfile="ddo44_fig1.ps" angle=0 vscale=90
    %hscale=90 voffset=-685 hoffset=-0}}\par
%\newpage
%\section{a}
\vspace{16.0cm}
 \caption{$R$ image of DDO 44 obtained with the 6-m telescope with a seeing
	of $0\farcs9$. The field size is 170$\arcsec\times178\arcsec$.
	North is up, East is left}
\end{figure*}

   In the present paper we study the basic properties and the environmental 
status of another nearby
dSph galaxy, DDO~44 (van den Bergh 1959). As a large-scale
photograph taken in the $B$ band (Karachentseva et al.\ 1985) shows,
DDO~44 = UGCA~133 (Nilson 1974) = kk\,61 (Karachentseva \& Karachentsev 1998)
has a very low surface brightness distribution without any visible knots
that might be indicative of H {\sc ii} regions or stellar associations.  
With an angular dimension of $3\farcm0\times2\farcm0$ the galaxy's integrated
apparent magnitude corresponds to $B_T= 15\fm64$ (NASA's Extragalactic 
Database =
 NED). In the NED, DDO~44 is classified as morphological type Sm, while van den Bergh
(1959) and Karachentseva \& Sharina classified it as dSph. Being only 80 arcmin
away from the bright Sc galaxy NGC~2403, DDO~44 seems to be a likely
companion of this galaxy. Unfortunately, the radial velocity of
DDO~44 is still unknown. Recent observation in the H {\sc i} line yields only
an upper limit for its H {\sc i} flux, $S < 6$ mJy (Huchtmeier et al.\ 1999).
Therefore a direct measurement of distance to DDO~44 can solve the question
of its physical connection to NGC~2403.

\section{Ground based observations and data reduction}
\begin{figure*}[bt]
%\vbox{
%\special{psfile="ddo44_fig2.ps" angle=0 vscale=91.5
   %hscale=91.5 voffset=-700 hoffset=-0}}\par
\vspace{16.0cm}
\caption{WFPC2 image of DDO~44 made from the combination of two 600s
	exposures through the F606W and F814W filters. A globular cluster
	candidate is indicated by the arrow in the left bottom corner.}
\end{figure*}
DDO~44 was observed on January 19, 1999 with a CCD camera of the SAO 6-meter
telescope. A CCD chip of 1024$\times$1024 pixels provided a total field of
$3\farcm5\times3\farcm5$ with a resolution of $0\farcs206$/pixel.

Two frames in the Cousins $R$ band with exposures of $2\times600$~s and four frames in $I$
band with exposures of $4\times300$~s were obtained. The seeing was
FWHM = $0\farcs9$. The $R$ band frame of DDO~44 is shown in Figure 1,
which shows the galaxy to be resolved into a large number of faint stars.
Absolute calibration and atmospheric extinction are based on observations of
twenty Landolt (1992) standard stars. We have obtained the following 
calibrations
(standard deviations of the residuals are 0.04 mag in $R$ and 0.05 mag in $I$,
respectively):
$$R-r = 27\fm38 + 0.0187\cdot(R-I)$$
$$I-i = 26\fm90 + 0.0513\cdot(R-I),$$
\noindent where $r,i$ are instrumental magnitudes corrected for atmospheric 
extinction.

 The derived images were processed with the MIDAS implementation of the DAOPHOT
II program (Stetson et al.\ 1990). A total of about 1000 stars were measured
to a limiting magnitude $I\sim24\fm2$. The rms uncertainty of aperture
corrections is $0\fm04$ in $R$ and $0\fm05$ in $I$. Together with the rms 
errors of the \{$r,i$\}-to-\{$R,I$\}  transformation  and uncertainties of the 
extinction corrections, 0\fm04 ($R$), 0\fm05 ($I$), these errors yield
0.07 mag zero-point rms uncertainty in the $R$ band, and 0.08 mag in the $I$ 
band.
The final color-magnitude diagram is presented in Fig.\ 3. Apart from stellar
photometry we also carried out aperture photometry of the galaxy in circular
diaphragms, which allow to measure the integrated magnitudes in both 
filters.
\begin{figure*}[bt]
\vbox{
\includegraphics{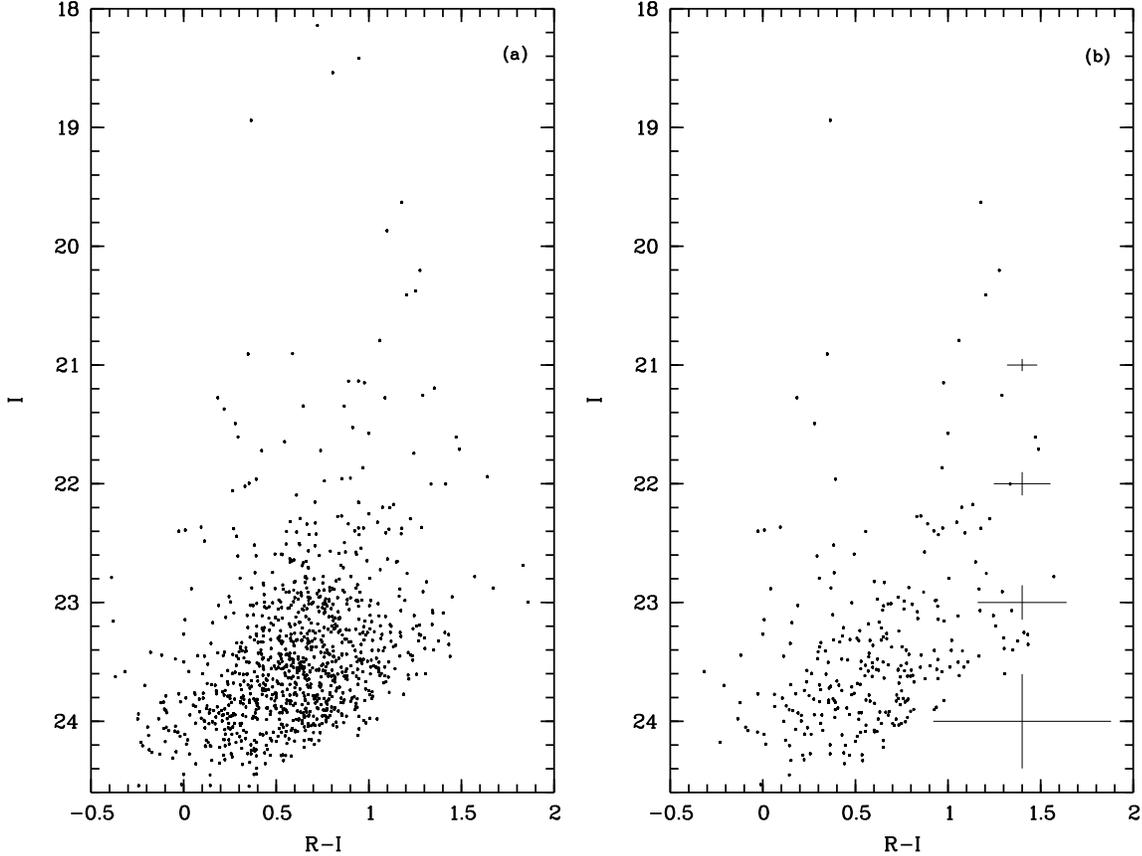}}\par
\vspace{14.0cm}
\caption{The color-magnitude diagram of resolved stars in the DDO~44 field.
  Stars in the inner ($R<75\arcsec$) region of DDO~44 are shown in the left
panel, while the stars in the surrounding field are presented in the right
panel. Errorbars in the right panel represent the standard errors derived
from photometry of artifical stars with different $I$ magnitudes and colors
$R-I$ within [0,1].}
\end{figure*}

\section{HST WFPC2 photometry}

  To study the resolved stellar population of the galaxy and to measure its
distance we observed DDO~44 = kk~61 with the Wide Field and Planetary Camera 2
(WFPC2) camera of the Hubble Space Telescope (HST). The galaxy
was observed on August 13, 1999 as part of a snapshot survey 
(program GO 8192, PI: Seitzer) to image 
probable nearby dwarf galaxy candidates from the list of Karachentseva \& 
Karachentsev (1998). Exposure times were 600 s in 
the F606W ($V$) and the F814W ($I$) filters, respectively.
Fig.~2 shows an image
of DDO~44 (both filters combined). The galaxy was centered on the WF3 chip.
After removing cosmic ray hits we carried out point source
photometry with the DAOPHOT II package in MIDAS. A total of about 4200 stars
were measured in both filters with an aperture radius of 1.5 pixels.
Then we determined the aperture correction from the 1.5 pixel radius 
aperture to the standard $0\farcs5$ radius aperture size for the WFPC2 
photometric system using bright uncrowded stars. Transformation of the 
instrumental magnitudes to the groundbased $V$, $I$ system
followed the prescriptions of Holtzman et al.\ (1995)
taking into account different relations for blue and red stars separately.
Objects with goodness of fit parameters 
$\mid$SHARP$\mid$ $> 0.3$, $\mid$CHI$\mid >$ 2, and
$\sigma (V) > 0.2$ mag were excluded. The final CMD in $I$, $V-I$ for 
1290 stars is shown in Fig.~4.

\section{Color-magnitude diagrams}
As mentioned above, the groundbased $I$, $R-I$ color-magnitude diagram 
(CMD) of DDO~44 is shown in Fig.~3.
The stars in the central part of the galaxy (with $r < 75\arcsec$)
are indicated in the left panel, and the stars at greater angular distances
are shown in the right. As is apparent from the CMD
the brightest stars with $I < 21$ mag are foreground stars. The central
zone of DDO~44 contains almost no stars bluer than $(R-I < 0\fm2$) with the
possible exception of the faintest stars, whose
colors may be affected by photometric errors. The error bars in the panel
of Fig.\ 3 indicate the standard errors of magnitude and color calculated
from artificial star experiments with $I = 21, 22, 23$, and 24 mag and
$0 < R-I < 1$ (ADDSTAR routine).

\begin{figure*}[bt]
\vbox{
\includegraphics{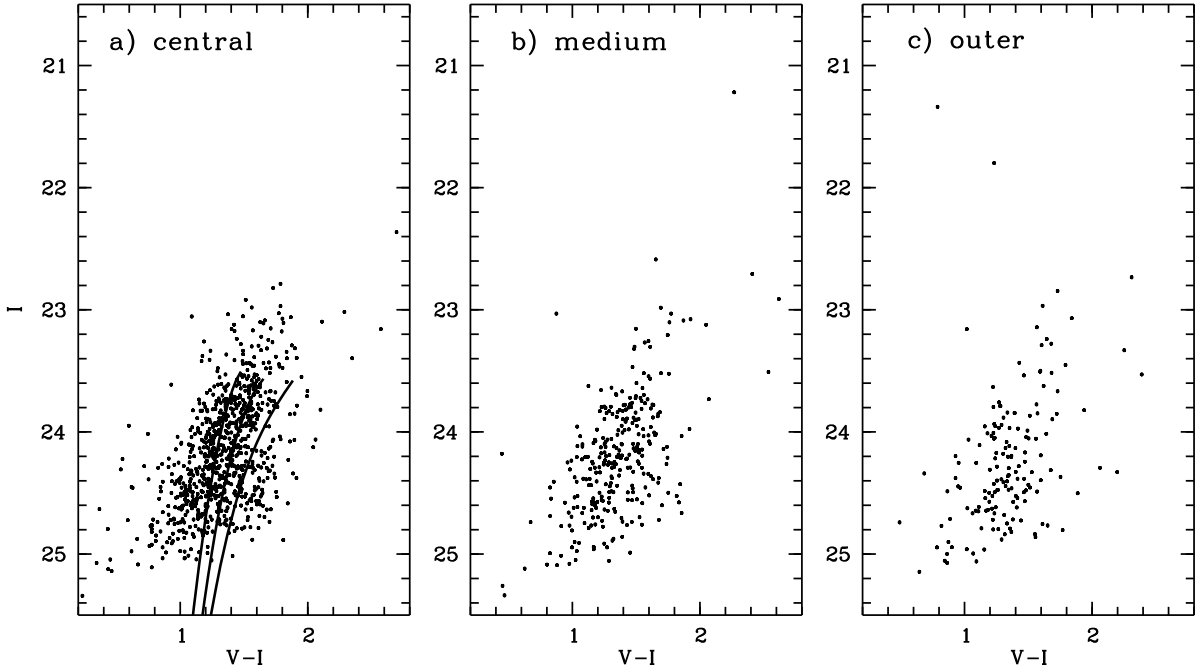}}\par
\vspace{13.0cm}
\caption{Color-magnitude diagram from WFPC2 data of DDO~44. The three panels
show diagrams based on stars at the central (WF3) field, medium and outer
field of equal $800 \times 800$ pixels area each. The solid lines in the 
left panel
show the loci of the RGBs of globular clusters with different metallicities:
M~15, M~2, and NGC~1851 from left to right.}
\end{figure*}
The most remarkable feature of the CMD is a
rather sharp rise in a number of stars fainter than $I = 22\fm8$ and 
predominantly
red colors $(R-I) > 0\fm5$. {From} the groundbased image alone one would 
tend to identify this feature with the red giant branch
(RGB), which yields a wrong distance to the galaxy.

   Fig.~4 shows CMDs of DDO~44 for the central WF3 field (left panel),
for the adjoining region covering the lower half of the WF2 chip and the 
left half of the WF4 chip in Fig.\ 2 (central panel in Fig.\ 4), and 
DDO~44's outer parts (the remaining halves of WF2 and WF4; right panel
of Fig.\ 4). In the central field the number of stars rises abruptly
at $I = 23\fm5$, which we interpret as the tip of the red giant branch (TRGB).
No bright blue stars with $V-I < 0\fm7$ are present. This allows us to 
estimate a lower limit for the most recent star formation episode in 
DDO~44.  Using the Bertelli et al.\
(1994) isochrones and adopting the DDO~44 distance and mean abundance
from the next sections, the absence of bright blue stars indicates that 
no star formation has occurred
in the galaxy for at least the last 300~Myr. 

   A significant number of red stars with  $22.8 < I < 23.5$ is evident 
in the left panels of Fig.~3 and Fig.~4.  These stars
are probably upper asymptotic giant branch (AGB) stars. When 
confused with the RGB stars, the clump of AGB stars can lead to a significant
underestimate of the galaxy distance in a shallow CMD.
The bona-fide AGB stars are indicative of a significant intermediate-age
population in DDO~44.  

\begin{figure}[bt]
\vbox{
\includegraphics{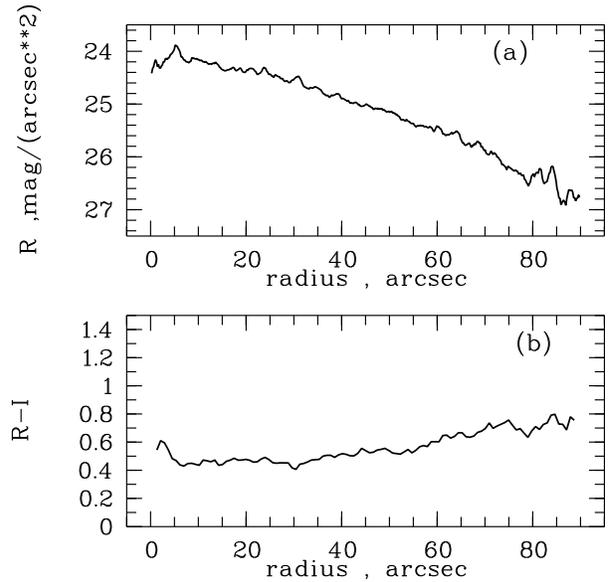}}\par
\vspace{9.0cm}
\caption
{Radial distribution of surface brightness (a) and
color (b) in DDO~44, averaged within circular annuli.}
\end{figure}

  According to Da Costa~\& Armandroff (1990), the tip of RGB can be assumed 
to be at $M_I = -4.05$ for metal poor systems. We find the apparent magnitude 
of the TRGB to be $I$(TRGB) = $23\fm55\pm0\fm15$, which yields a distance 
modulus of $(m-M)_0= 27\fm52\pm0\fm15$ or $D = (3.2\pm0.2)$ Mpc with a 
Galactic extinction
$A_I = 0\fm08$ from Schlegel et al.\ (1998). The solid lines in Fig.\ 4a are
globular cluster fiducials from Da Costa~\& Armandroff, which were reddened and
shifted to the galaxy distance.  The fiducials cover a range of [Fe/H] values
(from left to right): $-2.2$ dex (M~15), $-1.6$ dex (M~2), and $-1.2$ dex 
(NGC~1851).

\section{Metal abundance}

With the distance modulus of DDO~44 we can estimate the mean metallicity of
DDO~44 from the mean color of the RGB measured at an absolute magnitude
$M_I= -3\fm5$, as recommended by Da Costa~\& Armandroff (1990). Based on a
Gaussian fit to the color distribution of the giant stars in the range
$23\fm8 < I < 24\fm3$ we derive a mean dereddened color of the RGB stars of
$(V-I)_{0,-3.5} = 1\fm34\pm0\fm05$.  Following Lee et al.\ (1993)
this yields a mean metallicity  [Fe/H] = ($-1.7\pm0.4$) dex. Here the error
includes also contributions from the calibration relations (0.08~mag), and
a reddening uncertainty (0.03~mag).

\section{Integrated properties of DDO~44}

The radial distribution of surface brightness in the $R$ band frame (Fig.\ 1)
averaged over azimuth is shown in the upper panel of Fig.\ 5. The lower panel
reproduces the radial variation of the $(R-I)$ color also averaged in azimuth.
In the main body of DDO~44 $(10\arcsec< R < 70\arcsec$) the surface brightness
profile was approximated
by an exponential fit with a scale length $h = 39\arcsec\pm3\arcsec$. The
observed surface brightness is 
$\mu_R(0) = 24\fm1\pm0\fm2$~mag~arcsec$^{-2}$.
The mean galaxy color increases slightly towards the periphery of the galaxy, 
which may be caused by a radial age gradient in DDO~44.
The galaxy's total color index, $(R-I)_T = 0\fm51\pm0\fm05$, was determined as
the difference of asymptotic integral magnitudes in each band derived from
curves of growth. The sky background was calculated to be $R_{\rm sky} =
20\fm47\pm0\fm01$ and $(R-I)_{\rm sky} = 1\fm62\pm0\fm01.$
\begin{table}
\caption{Properties of DDO 44}
\begin{tabular}{lc} \hline
    Parameter  &                      DDO 44       \\
\hline
  RA  (1950.0)              &    07 29 13.1           \\
  Dec.(1950.0)              &    66 59 40              \\
  Galactic $l$                &    149.09                 \\
  Galactic $b$                &     28.96                  \\
  Dimension ($\arcmin$)         &            3.0$\times$2.0        \\
  $B_T$                   &              15.64            \\
  $E(B-V)$                &               0.04              \\
  Extinction: $A_B, A_I$  &            0.19, 0.08             \\
  R$_T$                   &             14.5$\pm$0.1               \\
  $(R-I)_T$               &             0.51$\pm$0.05                \\
  $\mu_R(0)$   (mag/($\arcsec)^2$)  &           24.1$\pm$0.2               \\
  I$_{TRGB}$                 &           23.55$\pm$0.15                   \\
  $(V-I)_{0,-3.5}$            &           1.34$\pm$0.05         \\
  ${\rm [Fe/H]}$                  &           $-1.7\pm0.4$            \\
  $(m-M)_0$                 &           27.52$\pm$0.15              \\
  $D_{MW}$  (Mpc)             &            3.2$\pm$0.2                \\
  $M^o_R$                   &            $-13.1$                      \\
  Linear diameter  (kpc)  &              2.8                         \\
  Scale length ($\arcsec$)        &             $39\pm3$              \\
  Scale length (kpc)      &             $0.60\pm0.04$         \\
  Type                    &             dSph           \\
  $S_N$                   &  6: \\
  Projected separation from NGC 2403 (kpc) &   75      \\
  Radial distance to NGC 2403 (kpc)  & $30\pm450$   \\            \hline

\end{tabular}
\end{table}
  A summary of the basic parameters of DDO~44 is given in Table 1. The data 
in the first six lines are from NED, while the other listed parameters are
from this paper. The symmetric shape of DDO~44, its smooth surface brightness
profile, the reddish total colors $(R-I) = 0\fm51$,  $(B-R) = 1\fm14$, and 
the lack of
an appreciable amount of neutral hydrogen favour DDO~44 classification as
a dwarf spheroidal system. The integrated absolute magnitude of DDO~44, $M_R =
-13\fm1$, and its linear diameter, 2.8 kpc, correspond to the parameters typical
for spheroidal companions of the Milky Way and M31. The derived parameters of
DDO~44 follow the general [Fe/H] vs.\ central surface brightness and [Fe/H]
vs.\ $M_V$ relationships defined by Local Group dwarf galaxies (Caldwell et 
al.\ 1998, Grebel~\& Guhathakurta 1999).

\begin{figure}[bt]
\vbox{
\includegraphics{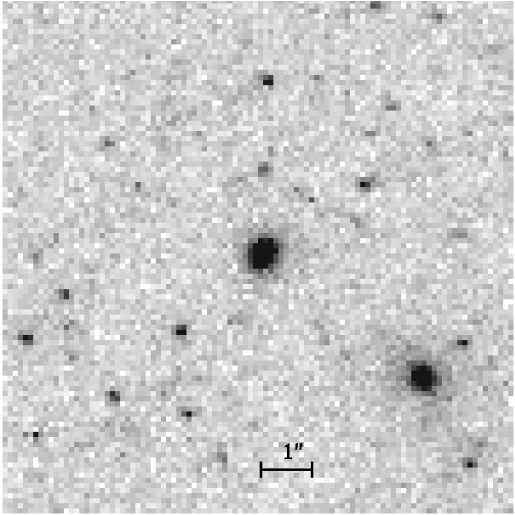}}\par
\vspace{9.0cm}
\caption
{Globular cluster candidate, shown in the F606W + F814W filters
from HST.}

\end{figure}

\section{Globular clusters}
The two brightest dSph galaxies in the Local Group, Fornax and
Sagittarius, each contain several globular clusters. In galaxies with 
$M_V = -13$ to $-14$ one expects to find 1--2 globulars on average
(Caldwell et al.\ 1998). Thus one may expect about one globular cluster
in DDO~44, which has $M_R= -13.1$. We searched for globular clusters
in DDO~44 and found one candidate with appropriate color, $V-I = 0\fm83$, and
magnitude, $V= 22\fm18$. Its position in the WF3 chip (Fig.\ 2) is indicated by
the arrow. An enlargement of the HST image of this candidate is presented in
Fig.\ 6.  Note that the globular cluster candidate is not found in the 
center of DDO~44; i.e., DDO~44 is not a nucleated dSph.
Another diffuse, but red object is situated in the right bottom corner. 

   The absolute magnitude of our globular cluster candidate, $M_V= -5\fm4$, 
and its dereddened color, $(V-I)_0= 0\fm77$, are quite similar to those of  
the metal-poor
Galactic globular cluster NGC~4147 (Peterson 1993). Fig.\ 7 presents the radial
profile of our globular cluster candidate, along with that of a star of about
the same brightness and color. From this profile the half-light radius of the
cluster candidate is $0\farcs27$ or 4.2~pc, well within the range of 2--10~pc
typical for half-light radii of Galactic globular clusters (Harris 1996).

\begin{figure}[bt]
\vbox{
\includegraphics{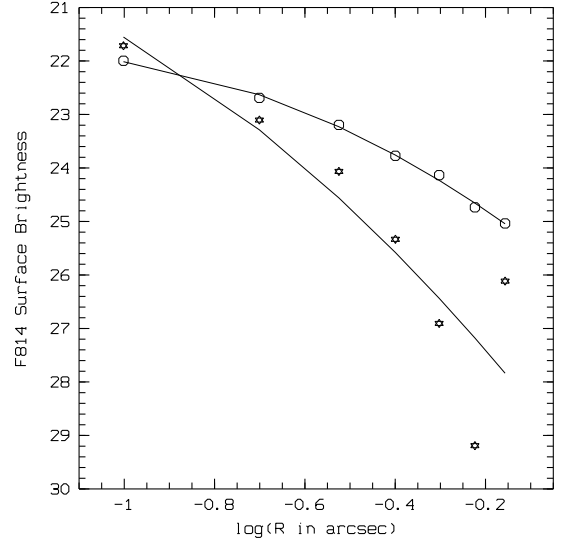}}\par
\vspace{9.0cm}
\caption
{Radial $I$ light profile of the globular cluster candidate (open
circles). The diamonds show the radial profile of a star of the same
magnitude.}
\end{figure}
   Assuming that our candidate is indeed a globular cluster, the resulting
specific frequency is $S_N \approx 6$.  The specific frequency is the number of
globular clusters normalized to a galaxy with $M_V = -15$, 
$S_N = N_{GC} \cdot 10^{0.4(M_V+15)}$ (Harris \& van den Bergh 1981), where 
$N_{GC}$ is the number of globular clusters
and $M_V$ the absolute $V$ magnitude of the dSph.  Our estimated $S_N$ is 
lower than that of Fornax and Sagittarius ($S_N > 10$), higher than the
mean specific frequency in  
non-nucleated dSphs and dwarf ellipticals ($S_N = 3.1\pm0.5$), and similar to 
that of nucleated dwarf ellipticals ($S_N = 6.5\pm1.2$; Miller et al.\ 1998). 

\section{Conclusions}

Based on deep $R$, $I$ and $V$, $I$ CCD images obtained with groundbased and 
space telescopes, we resolved the very low surface
brightness dwarf galaxy DDO~44 into stars for the first time. The CMDs for
$\sim$1290 stars show the presence of a red giant branch with 
$I$(TRGB) = $23\fm55\pm0\fm15$, which yields a true distance modulus of 
$27\fm52\pm0\fm15$. The corresponding distance
of DDO~44 from the Milky Way is $D_{MW} = (3.2\pm0.2)$ Mpc.
In projection DDO~44 lies only 80 arcmin or 75~kpc away from
NGC~2403 and, taken at face value, may be its companion. The distance modulus
of the spiral galaxy NGC~2403 derived from light curves of cepheids is
$27\fm5\pm0\fm3$ or $D = 3.2$ Mpc (Freedman 1988). Hence, both the galaxies
seem indeed to be located at a very similar distance.  

The large number of bona-fide AGB stars indicates that DDO~44 may have a
significant intermediate-age population.  This is also observed in the 
more distant dSph companions of the Milky Way.  

  A special search for LSB dwarf galaxies (Karachentseva~\& Karachentsev 1998)
did not reveal other dSph candidates in the extended surroundings of NGC~2403.
But two well-known irregular galaxies, NGC~2366 and K~52, are separated
3.7$\degr$ and 5.6$\degr$ from NGC~2403, respectively. Both of them have radial
velocities within 15~km/s of that of NGC~2403, and their linear projected
separations are 210~kpc and 310~kpc. Thus the small sample of companions
of NGC~2403 seems to exibit the same effect of morphological segregation as 
other nearby groups.

  As was noted by Karachentsev (1996), among the well-studied groups
around nearby giant galaxies the number of dSph companions
depends strongly on the morphological type of the main galaxy. In fact,
spheroidal systems have been detected only in the vicinity of E--Sb type
galaxies with massive bulges such as the
Milky Way (N$_{\rm sph}$ = 9), M~31 (6), M~81 (8), and NGC~5128 (7), but not
around sufficiently massive Sc galaxies like M~101, NGC~5236, M~33.
This morphological relation has probably an evolutionary significance. But
a presence of spheroidal dwarf companion arround NGC~2403 may be 
the first case of disagreement with the mentioned tendency.

\acknowledgements
{IK and MS thank N.\ Tikhonov, S.\ Kajsin, and I.\ Drozdowsky, who have 
taken part in the observations. This work was partially supported by INTAS-RFBR 
grant 95-IN-RU-1390. EG, DG, PG, PH, AS, and PS acknowledge support  by NASA through grant number GO-08192.97A from
the Space Telescope Science Institute, which is operated by the Association
of Universities for Research in Astronomy, Inc., under NASA contract
NAS5-26555.  EK acknowledges support by NASA through grant 
HF-01108.01-98A from the Space Telescope Science Institute.
This research has made use of the NASA/IPAC Extragalactic 
Database (NED) which is operated by the Jet Propulsion Laboratory, California 
Institute of Technology, under contract with NASA, and of
NASA's Astrophysics Data System Abstract Service.
Finally, we thank the referee, A.\ Aparicio, for suggestions that 
improved the presentation of the paper.}

{}
\newpage
\end{document}